\documentclass[11pt,twoside,a4paper,cmspaper,final,collab]{cms-tdr}
\def\svnVersion{0b7eb93}\def\svnDate{2026/10/18}\def\cmsCernNoTag{CERN-EP-2025-232}\def\cmsCernDate{\today}\def\cmsMessage{Published in Physical Review D as \href{http://dx.doi.org/10.1103/k9zs-c42w}{\doi{10.1103/k9zs-c42w}.}}
\begin{document}\cmsNoteHeader{SMP-24-002}

\newcommand{\sigeta}{\ensuremath{\sigma_{\eta\eta}}\xspace}
\newcommand{\Ich}{\ensuremath{I_{\text{ch}}}\xspace}
\newcommand{\OBW}{\ensuremath{\mathcal{O}_{\widetilde{B}W}}\xspace}
\newcommand{\OGP}{\ensuremath{\mathcal{O}_{G+}}\xspace}
\newcommand{\OGM}{\ensuremath{\mathcal{O}_{G-}}\xspace}
\newcommand{\mO}{\mathcal{O}}
\newcommand{\cut}{\Lambda}
\newcommand{\Zg}{\ensuremath{\PZ\PGg}\xspace}
\newcommand{\llg}{\ensuremath{\Pell\Pell\PGg}\xspace}
\newcommand{\signllg}{\ensuremath{\Pellp\Pellm\PGg}\xspace}
\newcommand{\mll}{\ensuremath{m_{\Pellp\Pellm}}\xspace}
\newcommand{\SUtwoL}{\ensuremath{\mathrm{SU}(2)_\mathrm{L}}\xspace}
\newcommand{\UoneY}{\ensuremath{\mathrm{U}(1)_{\mathrm{Y}}}\xspace}
\newcommand{\hthreeZ}{\ensuremath{h_3^\PZ}\xspace}
\newcommand{\hthreeG}{\ensuremath{h_3^\PGg}\xspace}
\newcommand{\hfourZ}{\ensuremath{h_4^\PZ}\xspace}
\newcommand{\hfourG}{\ensuremath{h_4^\PGg}\xspace}
\newcommand{\cGplus}{\ensuremath{\tilde{c}_{G+}}\xspace}
\newcommand{\cGminus}{\ensuremath{\tilde{c}_{G-}}\xspace}
\newcommand{\cBW}{\ensuremath{\tilde{c}_{\widetilde{B}W}}\xspace}
\newcommand{\UoneEM}{\ensuremath{\mathrm{U}(1)_{\mathrm{EM}}}\xspace}
\ifthenelse{\boolean{cms@external}}{\providecommand{\cmsLeft}{upper\xspace}}{\providecommand{\cmsLeft}{left\xspace}}
\ifthenelse{\boolean{cms@external}}{\providecommand{\cmsRight}{lower\xspace}}{\providecommand{\cmsRight}{right\xspace}}

\cmsNoteHeader{SMP-24-002}
\title{Measurement of \texorpdfstring{\Zg}{Zgamma} production in proton-proton collisions at  \texorpdfstring{$\sqrt{s}=13.6\TeV$}{sqrt(s) = 13.6 TeV} and constraints on neutral triple gauge couplings}

\date{\today}

\abstract{
A measurement of the \Zg production cross section in proton-proton collisions at a center-of-mass energy of 13.6\TeV is presented. Data corresponding to an integrated luminosity of $34.8\fbinv$, collected by the CMS experiment at the LHC in 2022 are used. Events with an oppositely charged pair of muons or electrons, with an invariant mass corresponding to a \PZ boson, together with an isolated photon are selected. The measured fiducial cross section for the combined electron and muon channels is $1.896 \pm 0.033 \stat \pm 0.054 \syst \pm 0.006 \thy\unit{pb}$, in agreement with the standard model prediction of $1.922 \pm 0.094\unit{pb}$. Constraints on neutral triple gauge couplings generated by dimension-8 operators in a recently proposed effective field theory framework are determined for the first time.
}

\hypersetup{
pdfauthor={CMS Collaboration},
pdftitle={Zgamma cross section measurement at 13.6 TeV},
pdfsubject={CMS},
pdfkeywords={CMS, muti-boson, Run3}}

\maketitle 

Measurements of the production of multiple vector bosons (\PW, \PZ, \PGg) at a single interaction vertex are crucial for understanding the electroweak (EW) sector of the standard model (SM). In particular, the non-Abelian nature of the EW interaction predicts that self-interactions occur among the vector bosons, resulting in a wide range of multiboson production mechanisms. 

\begin{figure*}[htbp]
\centering
\includegraphics[width=0.9\textwidth]{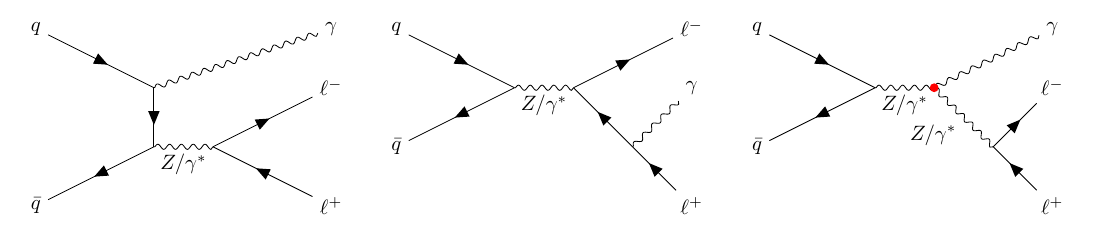}
\topcaption{
Leading order Feynman diagrams for \Zg production in proton-proton collisions and an example anomalous gauge coupling diagram. Left: initial-state radiation. Center: final-state radiation. Right: diagram involving anomalous gauge couplings that are forbidden in the SM at tree level. The red point indicates the anomalous $\PZ\PGg\PZst$ or $\PZ\PGg\PGg^{\ast}$ gauge coupling vertices.}\label{INTP1}
\end{figure*}

The production of \Zg is one of the most precisely measurable diboson processes at the CERN LHC due to its large cross section and clean signature from the leptonic decays of the Z boson. Its observation in the $\PZ \to \Pellp \Pellm$ ($\Pell = \Pe$ or $\Pgm$) decay channel has little background, which allows precise measurements and tests of SM predictions. Illustrative leading order (LO) Feynman diagrams for \Zg production and an example of an anomalous gauge coupling diagram are shown in Fig.~\ref{INTP1}.

In addition, the \Zg process provides a powerful tool for placing constraints on neutral triple gauge couplings (nTGCs). These couplings do not appear among the dimension-4 terms of the
SM Lagrangian, and they are not generated by dimension-6
terms in SM extensions via the SM effective
field theory (SMEFT)~\cite{Buchmuller:1985jz,Grzadkowski:2010es,Ellis:2014dva}.
Instead, nTGCs first appear through the gauge invariant dimension-8 operators in the SMEFT~\cite{Gounaris:1999kf,Gounaris:2000dn,Degrande:2013kka,Ellis:2020ljj}. Hence any indication of nonvanishing nTGCs would provide direct evidence for physics beyond the SM (BSM), which is different in nature from anything that might be generated by dimension-6 operators. The high-mass
part of the \Zg invariant mass distribution is sensitive to nTGCs, allowing the detection of BSM physics. 

Measurements of \Zg production have been performed by experiments at the LEP~\cite{L3:2004hlr,OPAL:2000ijr}, the Tevatron~\cite{CDF:2011rqn}, and the LHC~\cite{ATLAS:2019gey,ATLAS:2016qjc,CMS:2016cbq,ATLAS:2022wnf} colliders. So far, BSM physics searches using \Zg production~\cite{ATLAS:2019gey,ATLAS:2016qjc,CMS:2016cbq} have relied on a vertex parameterization model (VPM), which only considers the unbroken $\mathrm{U}(1)_{\mathrm{EM}}$ gauge symmetry~\cite{Gounaris:2000tb,Degrande:2013kka}, and is in contrast to the dimension-8 operators in the SMEFT that preserve the full EW gauge group $\SUtwoL\otimes\UoneY$ of the SM. 
A new theoretical framework for nTGCs has recently been proposed~\cite{Ellis:2022zdw}, in which the nTGC interactions are derived from dimension-8 operators evaluated in the EW broken phase. The conventional approach of modifying the TGC vertex coupling constants leads to a violation of $\SUtwoL\otimes\UoneY$ gauge symmetry, whereas the new theoretical framework preserves the $\SUtwoL\otimes\UoneY$ gauge symmetry. We denote this dimension-8 gauge symmetry preserving nTGC model here as GSPM.

Comparing the nTGCs predicted by dimension-8 operators with those obtained from the vertex formulation, the authors of Ref.~\cite{Ellis:2022zdw} relate the four VPM parameters \hthreeZ, \hthreeG \hfourZ, and \hfourG to the coefficients of the dimension-8 GSPM operators $\mO_{G+}$, $\mO_{G-}$, and $\mO_{\widetilde{B}W}$, as follows:   
\begin{equation}\label{eq:h-dim8}
\begin{aligned}
\hfourZ &=-
\frac{\text{sign}(\cGplus)}{\Lambda^4_{G+}}
\frac{v^2 m_\PZ^2}{s_{W}c_{W}}, 
&& 
\hthreeZ, \hthreeG=0, 
&& \text{for }\OGP, 
\\
h_4^{\PGg} &=-
\frac{\text{sign}(\cGplus)}{\Lambda^4_{G+}}
\frac{v^{2}m_{\PZ}^{2}}{c_{W}^{2}}, 
&&
\hthreeZ, \hthreeG=0, 
&& \text{for }\OGP, 
\\
\hthreeZ &= 
\frac{\text{sign}(\cBW)}
{\Lambda_{\widetilde{B}W}^4}
\frac{v^{2}m_{\PZ}^{2}}{2s_{W}c_{W}},
&&
\hthreeG, \hfourZ, \hfourG=0,
&& \text{for }\OBW, 
\\
\hthreeG &= 
-\frac{\text{sign}(\cGminus)}
{\Lambda_{G-}^4}
\frac{v^{2}m_{\PZ}^{2}}{2c_{W}^{2}},
&&
\hthreeZ,\hfourZ,\hfourG=0,
&& \text{for }~\OGM. 
\end{aligned}
\end{equation}
Here $\text{sign}(\cGplus)$, $\text{sign}(\cGminus)$, and $\text{sign}(\cBW)$ are dimensionless coefficients that can take either sign and typically have a magnitude of 1, $m_\PZ$ is the mass of the Z boson, $v$ is the Higgs field's vacuum expectation value, $\cut_{G+}$, $\cut_{G-}$, and $\cut_{\widetilde{B}W}$ are the effective cutoff scales for BSM physics, and $s_W$ and $c_W$ are the sine and cosine of the weak mixing angle. 
 
In theoretical predictions~\cite{Ellis:2022zdw}, the difference between GSPM and VPM lies in $h_4^{\PZ}$ and $h_4^{\PGg}$. This can be seen from the $\PZ\PGg(\PZ,\PGg)^{\ast}$ vertex. For VPM, it is,
\ifthenelse{\boolean{cms@external}}{
\begin{multline}
\label{eq:FF0-nTGC}
\Gamma_{\PZ\PGg(\PZ,\PGg)^\ast}^{\alpha\beta\mu}
(q_1,q_2,q_3)
=
\frac{e\left(q_3^2 - m_{\PZ,\PGg}^2\right)}{m_\PZ^{2}}
\\\left(
   h_3^{\PZ,\PGg} q_{2\nu}\epsilon^{\alpha\beta\mu\nu}
   +
   \frac{h_4^{\PZ,\PGg}}{m_\PZ^2}
   q_2^{\alpha}q_{3\nu}q_{2\sigma}
   \epsilon^{\beta\mu\nu\sigma}\right),
\end{multline}}{
\begin{equation}
\label{eq:FF0-nTGC}
\Gamma_{\PZ\PGg(\PZ,\PGg)^\ast}^{\alpha\beta\mu}
(q_1,q_2,q_3)
=
\frac{e\left(q_3^2 - m_{\PZ,\PGg}^2\right)}{m_\PZ^{2}}
\left(
   h_3^{\PZ,\PGg} q_{2\nu}\epsilon^{\alpha\beta\mu\nu}
   +
   \frac{h_4^{\PZ,\PGg}}{m_\PZ^2}
   q_2^{\alpha}q_{3\nu}q_{2\sigma}
   \epsilon^{\beta\mu\nu\sigma}\right),
\end{equation}
}
where $q_1$, $q_2$, $q_3$ are the momenta of the gauge bosons, and $m_{\PZ,\PGg}$ is the mass of the \PZ boson or the photon. For GSPM, the $\PZ\PGg(\PZ,\PGg)^{\ast}$ vertex is, 
\ifthenelse{\boolean{cms@external}}{
\begin{multline}
\label{eq:ZAV*-FormF}
\Gamma_{\PZ\PGg(\PZ,\PGg)^{\ast}}^{\alpha\beta\mu(8)}
(q_1,q_2,q_3) =
\frac{e\left(q_3^2 - m_{\PZ,\PGg}^2\right)}{m_{\PZ}^2}
\\ \left[\left( h_3^{\PZ,\PGg} + h_4^{\PZ,\PGg}\frac{q_3^2}{2m_{\PZ}^2}\right)
q_{2\nu}\epsilon^{\alpha\beta\mu\nu}
 + \frac{h_4^{\PZ,\PGg}}{m_\PZ^2} 
q_2^{\alpha}q_{3\nu}q_{2\sigma}\epsilon^{\beta\mu\nu\sigma}\right].
\end{multline}
}{
\begin{equation}
\label{eq:ZAV*-FormF}
\Gamma_{\PZ\PGg(\PZ,\PGg)^{\ast}}^{\alpha\beta\mu(8)}
(q_1,q_2,q_3) =
\frac{e\left(q_3^2 - m_{\PZ,\PGg}^2\right)}{m_{\PZ}^2}
\left[\left( h_3^{\PZ,\PGg} + h_4^{\PZ,\PGg}\frac{q_3^2}{2m_{\PZ}^2}\right)
q_{2\nu}\epsilon^{\alpha\beta\mu\nu}
 + \frac{h_4^{\PZ,\PGg}}{m_\PZ^2} 
q_2^{\alpha}q_{3\nu}q_{2\sigma}\epsilon^{\beta\mu\nu\sigma}\right].
\end{equation}
} 
Since both \hfourZ and \hfourG correspond to the same operator \OGP, this leads to the relation $\hfourG = (s_W/c_W)h_4^{\PZ}$. In this Letter, we focus only on the three independent VPM parameters $\hthreeZ$, $\hthreeG$, and $\hfourZ$.

This Letter reports on a measurement of the \Zg production in the $\Pellp\Pellm\PGg$ decay channel with the CMS experiment using proton-proton ($\Pp\Pp$) collision data delivered by the LHC during 2022 at a center-of-mass energy of $\sqrt{s}=13.6\TeV$, with an integrated luminosity of 34.8\fbinv. Focus is placed on a fiducial cross section measurement and the determination of constraints on BSM contributions to this process that can modify the \Zg vertex. Constraints on nTGCs generated by GSPM are determined for the first time~\cite{Ellis:2022zdw}.

The CMS apparatus~\cite{CMS:2008xjf,CMS:2023gfb} is a multipurpose, nearly hermetic detector, designed to trigger on~\cite{CMS:2020cmk,CMS:2016ngn,CMS:2024aqx} and identify electrons, muons, photons, charged and neutral hadrons~\cite{CMS:2020uim,CMS:2018rym,CMS:2014pgm}. A global ``particle-flow'' (PF) algorithm~\cite{CMS:2017yfk} aims to reconstruct all individual particles in an event, combining information provided by the all-silicon inner tracker and by the crystal electromagnetic (ECAL) and brass-scintillator hadron calorimeters, operating inside a 3.8\unit{T} superconducting solenoid, with data from the gas-ionization muon detectors embedded in the flux-return yoke outside the solenoid. The reconstructed particles are used to build \PGt leptons, jets, and missing transverse momentum~\cite{CMS:2018jrd,CMS:2016lmd,CMS:2019ctu}.

Events of interest are selected using a two-tiered trigger system. The first level, composed of custom hardware processors, uses information from the calorimeters and muon detectors to select events at a rate of around 100\unit{kHz} within a fixed latency of about 4\mus~\cite{CMS:2020cmk}. The second level, known as the high-level trigger, consists of a farm of processors running a version of the full event reconstruction software optimized for fast processing and reduces the event rate to a few kHz before data storage~\cite{CMS:2016ngn,CMS:2024aqx}. This analysis uses a collection of high-level trigger paths for leptons, including
single and double lepton trigger paths, to ensure a trigger efficiency exceeding 99$\%$ for events that satisfy the subsequent offline selection criteria.

The primary vertex (PV) is taken to be the vertex corresponding to the hardest scattering in the event, evaluated using tracking information alone, as described in Section 9.4.1 of Ref.~\cite{CMS-TDR-15-02}.

Electrons and photons are measured in the pseudorapidity range of $\abs{\eta} < 2.5$, corresponding to the acceptance of the tracker. The energy of photons is obtained from the ECAL measurement. The energy of an electron is determined from a combination of the electron momentum at the primary interaction vertex as determined by the tracker, the energy of the corresponding ECAL cluster, and the energy sum of all bremsstrahlung photons spatially compatible with originating from the electron track. Identification criteria for photons include selections on the ratio of hadronic to electromagnetic energy deposited in calorimeters and a variable \sigeta quantifying the lateral extension of the shower~\cite{CMS:2020uim}. Large anomalous signals as well as other spurious and noise contributions are rejected by the photon reconstruction algorithm~\cite{CMS:2024ppo} using the techniques discussed in references~\cite{CMS:2015myp,CMS:2020xlg}.

Muons are measured in the range $\abs{\eta}<2.4$, with detection planes made using three technologies: drift tubes, cathode strip chambers, and resistive-plate chambers. The efficiency to reconstruct and identify muons is greater than 96\% over the full $\eta$ range. Matching of muons to tracks measured in the silicon tracker results in a relative transverse momentum \pt resolution of 1\% (3\%) in the barrel (endcaps), for muon \pt up to 100 GeV~\cite{CMS:2018rym}. 

The signal process $\Pp\Pp\to\Pellp\Pellm\PGg$ is simulated at next-to-leading order (NLO) precision in quantum chromodynamics (QCD) using \MGvATNLO version 2.9.13~\cite{Alwall:2014hca} with up to one jet in the matrix element calculation, merged with jets from the parton showering using the FxFx matching scheme~\cite{Frederix:2012ps}. Parton showering and hadronization are performed using \PYTHIAviii version 8.226~\cite{Sjostrand:2014zea}, with the CP5~\cite{Sirunyan:2019dfx} underlying event tune and NNPDF3.1~\cite{NNPDF:2017mvq} parton distribution functions (PDFs) at next-to-NLO (NNLO) accuracy. Background processes containing two prompt leptons and a prompt photon, including \ttbar process and $\PV\PV$ processes (where $\PV = \PW$ or $\PZ$), are simulated with different event generators. The \ttbar process is simulated  with \POWHEG at NLO in perturbative QCD and with \PYTHIAviii at LO, respectively. The corresponding cross section is normalized to a NNLO  prediction~\cite{CMS:2023qyl}. The $\PV\PV$ processes are simulated with \PYTHIAviii at LO. Finally, the detector response is simulated using the \GEANTfour~\cite{Agostinelli:2002hh} package, and the simulated events are reconstructed with the same reconstruction algorithms as used for the observed events.

Signal events must contain two same-flavor, oppositely charged leptons and a photon. Muons are required to pass a set of tight identification criteria, whereas electrons are required to meet the medium identification criteria~\cite{CMS:2018rym,CMS:2020uim}. All leptons are required to be isolated from the photon, quantified by an angular distance between the photon and the lepton momentum directions: $\Delta R_{\ell\gamma}=\sqrt{\smash[b]{(\Delta\eta)^2+(\Delta\phi)^2}}>0.5$, where $\phi$ is the azimuthal angle in radians. Muons should satisfy $\abs{\eta}<2.4$, while electrons should satisfy $\abs{\eta}<2.5$. Electrons falling within the range $1.44<\abs{\eta}<1.57$ are excluded, since the reconstruction of electromagnetic objects in this region is less effective. Leading (subleading) leptons, defined as the one with the highest (second-highest) \pt, should satisfy $\pt > 25\,(20) \GeV$.  To reduce the contribution from final-state radiation (FSR) and to limit $\llg$ production to a region dominated by on shell Z boson production, the dilepton invariant mass is required to satisfy $71 < \mll < 111 \GeV$. Photons must satisfy $\pt > 30\GeV$ and $\abs{\eta} < 2.5$, excluding the $1.44<\abs{\eta}<1.57$ range with the similar reason as mentioned above for electrons. To better distinguish between photons and electrons, photons should also have no corresponding track with a hit in the innermost part of the silicon tracker. These conditions together define the signal region.

The background due to events containing nonprompt photons, including genuine photons within jets, is estimated from the observed events. A two-dimensional sideband method is used to estimate this background in each bin of each distribution, following previous analyses~\cite{ATLAS:2022wnf,ATLAS:2019gey}. Two variables are used: the charged isolation sum denoted \Ich, and \sigeta~\cite{CMS:2020uim}. Independent control regions with a high purity in misidentified jets that pass the photon selection criteria are defined by inverting the two photon identification requirements individually and simultaneously, resulting in three control regions. This exploits the fact that misidentified jets tend to be less well isolated and have broader showers than prompt photons. For a given selection and observable bin, events in the inverted \sigeta region are used to predict the number of misidentified jet events in the signal region. For inverted regions with bins that are not populated, the nonprompt photon yields, and their statistical uncertainties are estimated by extrapolating from adjacent bins. A power-law function is fitted to the yields in the populated bins and used for the extrapolation in the nTGCs analysis. The associated systematic uncertainty is assessed by comparing the extrapolated results obtained using the power-law fit and an alternative exponential fit. The contribution of prompt photons originating from pileup interactions are estimated using Drell-Yan MC and applied as a correction factor to the data-driven photon background estimate.  They contribute approximately 4.1\% of the total observed yield in the signal region.

The fiducial phase space of the measurement is defined by a set of criteria imposed on signal events at generator level, and is used to calculate the fiducial cross section. Leptons at the generator level are defined at the so-called ``dressed level'', by combining their four-momenta with those of photons found within a cone of $\Delta R = 0.1$ around the lepton~\cite{CMS:2019raw}, to account for photon FSR. The fiducial region is defined by requiring the leading (subleading) lepton to have $\pt > 25\,(20)\GeV$  and $\abs{\eta} < 2.4$. The dilepton invariant mass must satisfy $71 < \mll < 111 \GeV$. For the generator-level  photon $\pt > 30\GeV$ and $\abs{\eta} < 2.5$ (excluding $1.44 < \abs{\eta} < 1.57$) are required, and an angular separation $\Delta R_{\Pell \PGg} > 0.5$ between the lepton and photon. Events that are reconstructed within the signal region but lie outside the fiducial phase space are categorized as out-of-fiducial events.

\begin{figure}[htbp]
\centering
\includegraphics[width=0.42\textwidth]{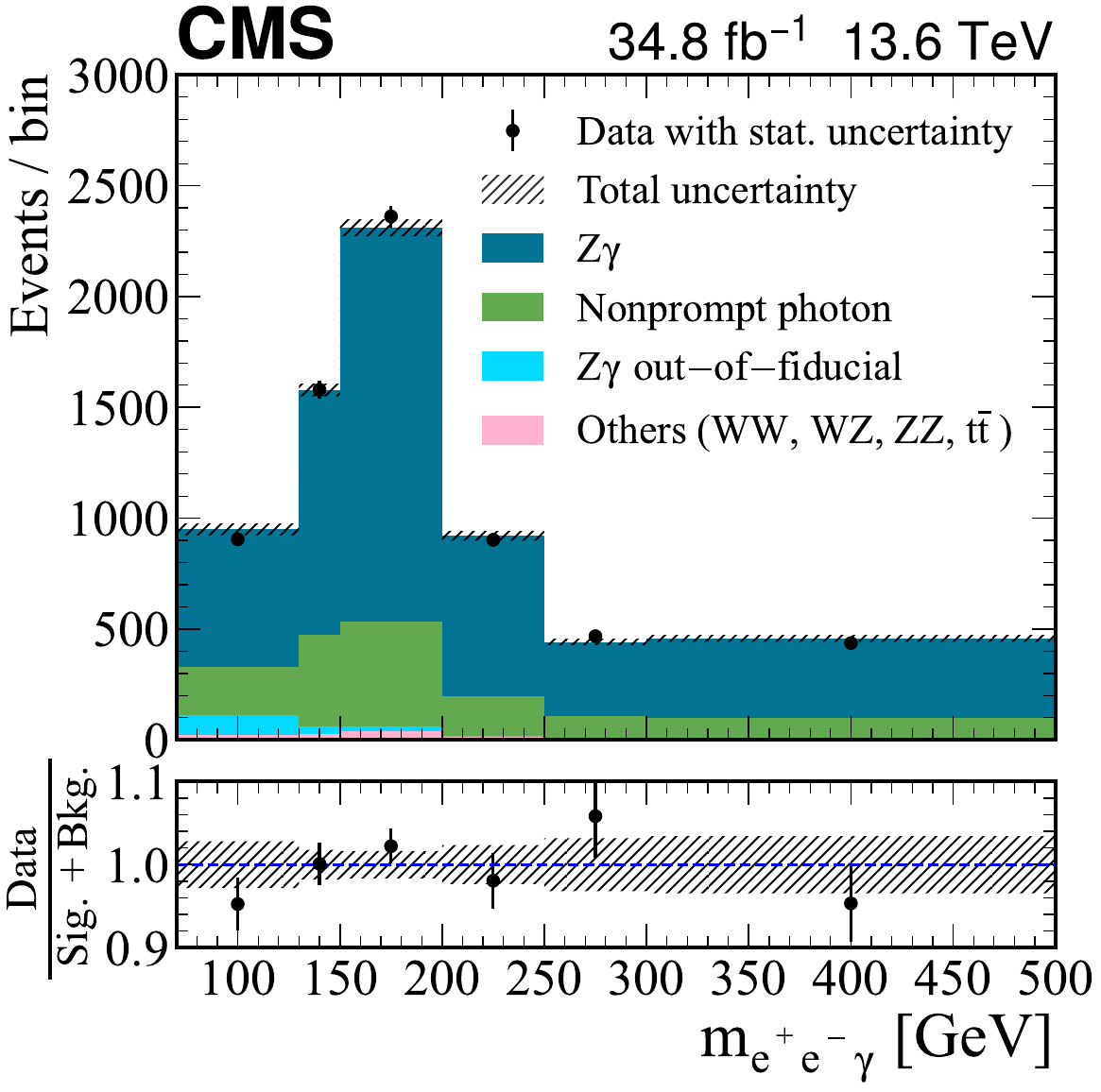}
\includegraphics[width=0.42\textwidth]{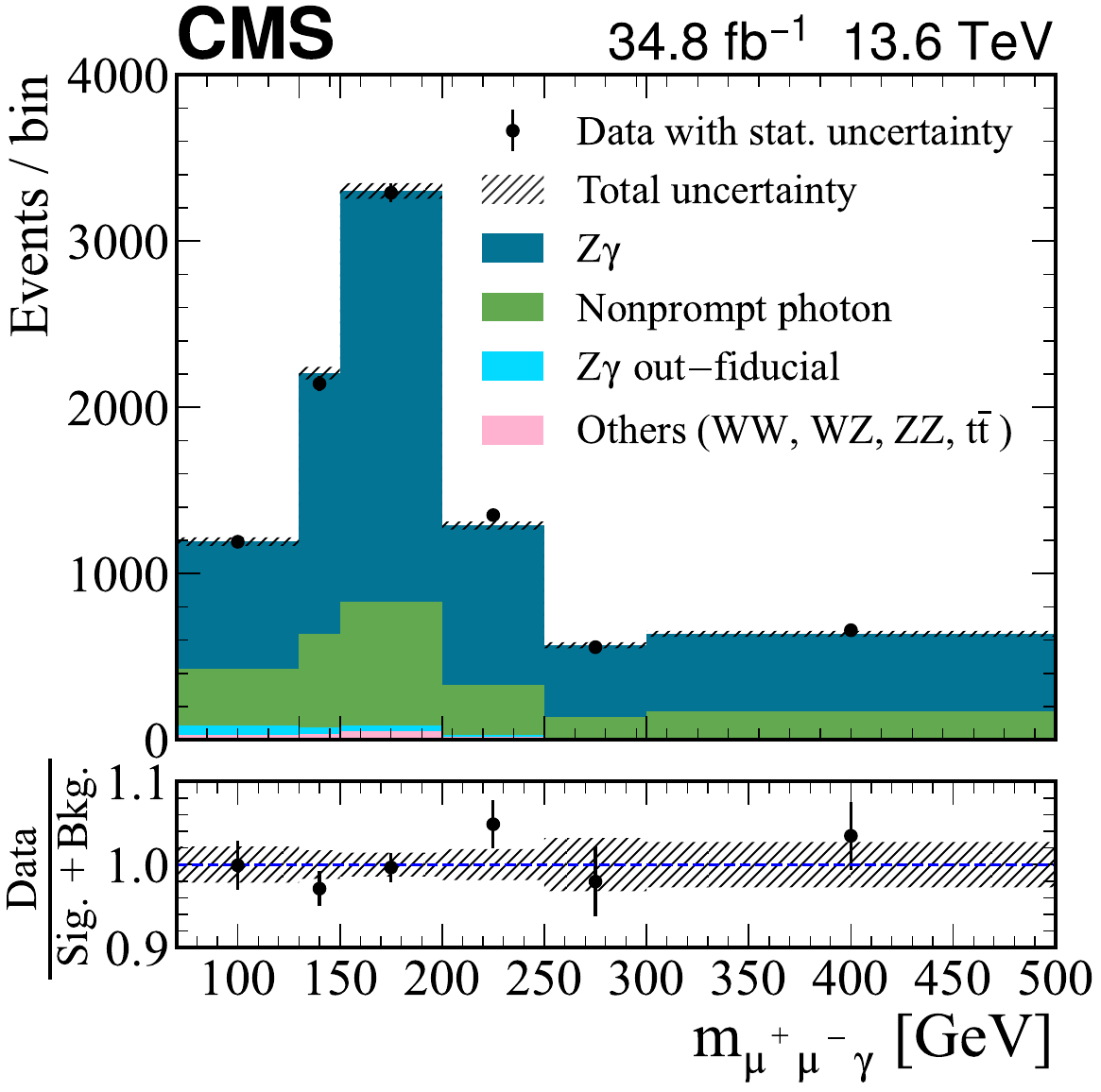}
\topcaption{Postfit distributions of $m_{\llg}$ in the electron (\cmsLeft) and muon channel (\cmsRight). The experimental data are shown as points with associated statistical uncertainties, while the hatched bands represent the total uncertainties. Uneven bin sizes GeV are used to ensure sufficient statistical precision in each bin. The last bins include overflow events.}\label{fig:datacards_bin}
\end{figure}

The signal strength (the ratio of the measured to predicted cross sections) is extracted from a binned maximum likelihood fit to the invariant mass of ${m_{\llg}}$ distribution, using the CMS statistical analysis tool \textsc{Combine}~\cite{CMS:2024onh}, based on the \textsc{RooFit}~\cite{Verkerke:2003ir} and \textsc{RooStats}~\cite{Moneta:2010pm} frameworks. The variable ${m_{\llg}}$ is computed as the invariant mass formed by the reconstructed four-momenta of the two leptons and the photon, and the likelihood function is the product of a Poisson probability density function for each ${m_{\llg}}$ bin. A simultaneous fit of the electron and muon channels is used for the results; in addition, electron- and muon-only fits are performed as consistency checks. The fit, shown in Fig.~\ref{fig:datacards_bin}, is performed in the ${m_{\llg}}$ range of 70--500\GeV in 6 bins. Uneven bin sizes $m_{\llg} = [70, 130,150, 200, 250, 300, 500]\GeV$ are used to ensure sufficient statistical precision in each bin.

Various sources of uncertainties are included in the fit as nuisance parameters subject to Gaussian constraints. Theoretical sources of uncertainty include the choice of the renormalization and factorization scales, PDFs, and parton shower modeling. These theoretical uncertainties are applied to signal and \ttbar processes. The two scales are varied by factors of 2 and 0.5 independently. The envelope of these variations, excluding the two extreme (2, 0.5) and (0.5, 2) cases, is taken as the uncertainty. The systematic uncertainty in the PDFs is calculated using the PDF4LHC15\_nnlo\_30\_pdf as PDF replicas, following the PDF4LHC group prescription~\cite{Butterworth:2015oua,Harland-Lang:2014zoa,NNPDF:2014otw,Dulat:2015mca}. Parton shower modeling uncertainties, which arise from the renormalization of QCD-induced initial-state radiation (ISR) and FSR, are the dominant theoretical uncertainty in the measurement. Variations of the ISR and FSR contributions consist of 4 combinations made by keeping one constant and doubling or halving the other. The experimental sources of systematic uncertainties include uncertainties in the estimate of the nonprompt photon background from control samples in data, as well as those associated with lepton and photon identification efficiencies, the lepton trigger, pileup~\cite{CMS:2016cbq} (additional inelastic pp collisions in the same or adjacent bunch crossings), where the uncertainty arises from the imperfect knowledge of the inelastic pp cross section, and integrated luminosity~\cite{CMS:2024sxz}. The statistical uncertainties from the limited size of the simulated signal and background samples are also included in the final fit, using the Barlow--Beeston-lite approach~\cite{Barlow:1993dm}. Among all types of uncertainties, the most significant contribution arises from the photon identification (1.67\%), followed by the integrated luminosity uncertainty (1.4\%). The total uncertainty on the fiducial cross section is 3.4\%.
  
The fiducial cross section for the $\signllg$ process is calculated as:
\begin{equation}
\sigma_{\text{fid}_{\signllg}} = \mu \sigma_{\text{gen\_fid}},
\end{equation}
where $\mu$ is the signal strength obtained from the $m_{\llg}$ fit and $\sigma_{\text{gen\_fid}}$ is the predicted cross section in the fiducial region from the generator, $\sigma_{\text{gen\_fid}} = 1.922 \pm 0.094\unit{pb}$, where the uncertainty is obtained from scale and PDF variations. The observed fiducial cross sections are shown in Table~\ref{tab:fiducial} for the electron and muon channels, and combined. The observed combined result $1.896 \pm 0.033 \stat \pm 0.054 \syst \pm 0.006 \thy\unit{pb}$ is in good agreement with the predictions, within uncertainties.

\begin{table}[htbp]
    \centering
    \topcaption{The observed fiducial $\signllg$ cross section of the process in each channel. The quoted uncertainties include statistical, systematical and theoretical components. Branching ratio $\mathcal{B}$ is for the $\PZ \to \signllg$ process.}
    \label{tab:fiducial}
    \begin{scotch}{c c} 
          Channel
          &
          $\sigma \mathcal{B} \pm \text{stat} \pm \text{syst} \pm \text{theo}\,(\mathrm{pb})$\\ 
        \hline 
         \Zg ($\PGm\PGm$) & $0.928 \pm 0.018 \pm 0.027 \pm 0.004$ \\  
         \Zg ($\Pe\Pe$) & $0.975 \pm 0.021 \pm 0.038 \pm 0.003$ \\  
         \Zg (combined) & $1.896 \pm 0.033 \pm 0.054 \pm 0.006$
    \end{scotch}
\end{table}

To set constraints on
nTGCs, VPM is implemented in \MGvATNLO with the universal \textsc{FeynRules} output interface~\cite{Degrande:2011ua}, while GSPM is implemented in \MGvATNLO directly as in Ref.~\cite{Ellis:2022zdw}. The nTGC sample generation takes advantage of the reweighting method in \MGvATNLO at LO. This method associates additional weights corresponding to other theoretical hypotheses within a sample, enabling the test of many nTGC parameter values using a single sample. For each of the nTGC operators, 17 weights are generated, including an nTGC coupling of zero representing the SM, 8 positive and 8 negative couplings. To ensure adequate number of events in the high-energy phase space, where the sensitivity to potential BSM physics is enhanced, a nonzero anomalous coupling close to the limits reported in previous publications is employed: $\hfourZ = -2.3824 \times 10^{-5}$ for VPM and $\cGplus = 0.02$ for GSPM.

\begin{figure}[htbp]
\centering
    \includegraphics[width=0.42\textwidth]{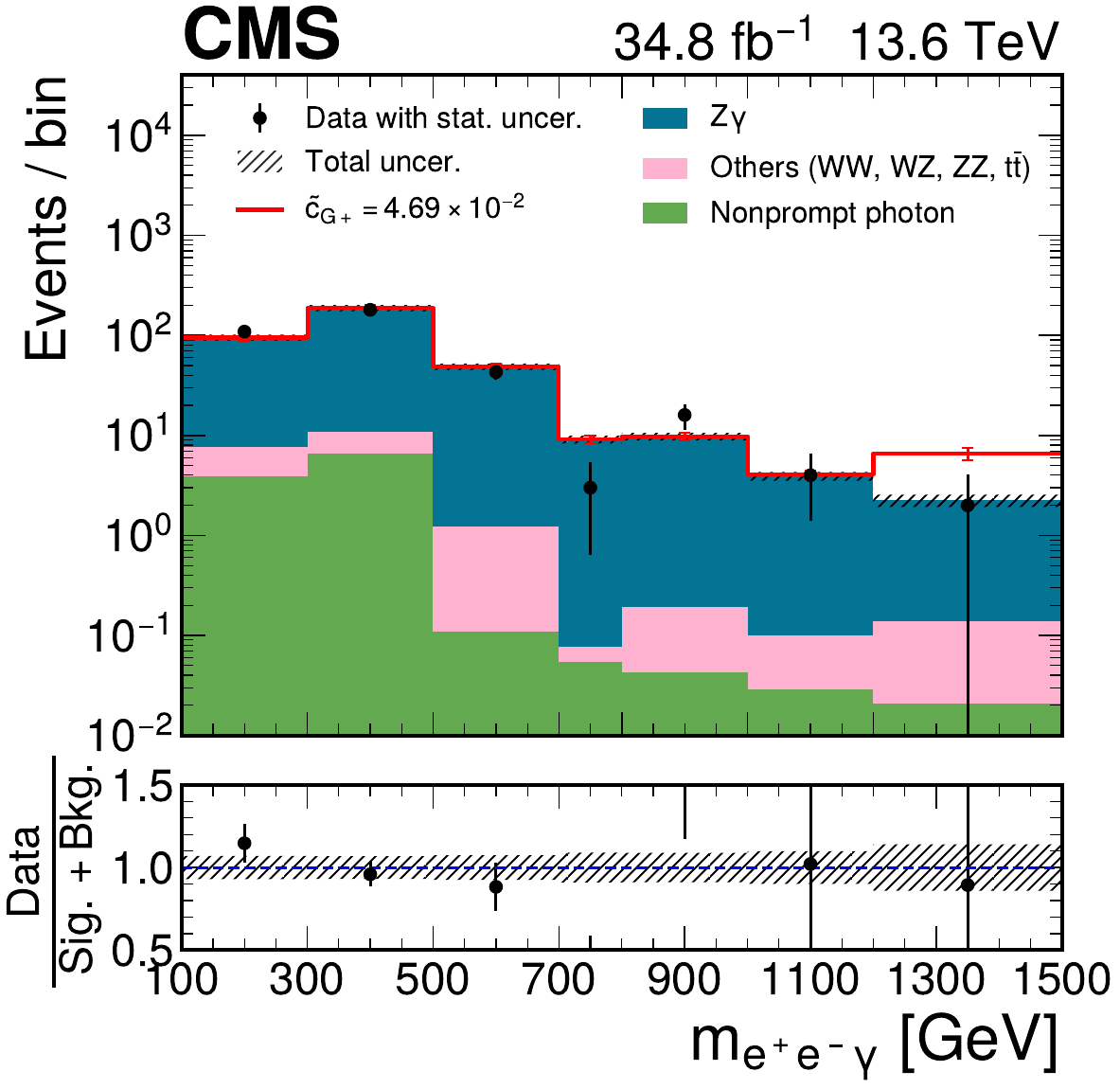}
    \includegraphics[width=0.42\textwidth]{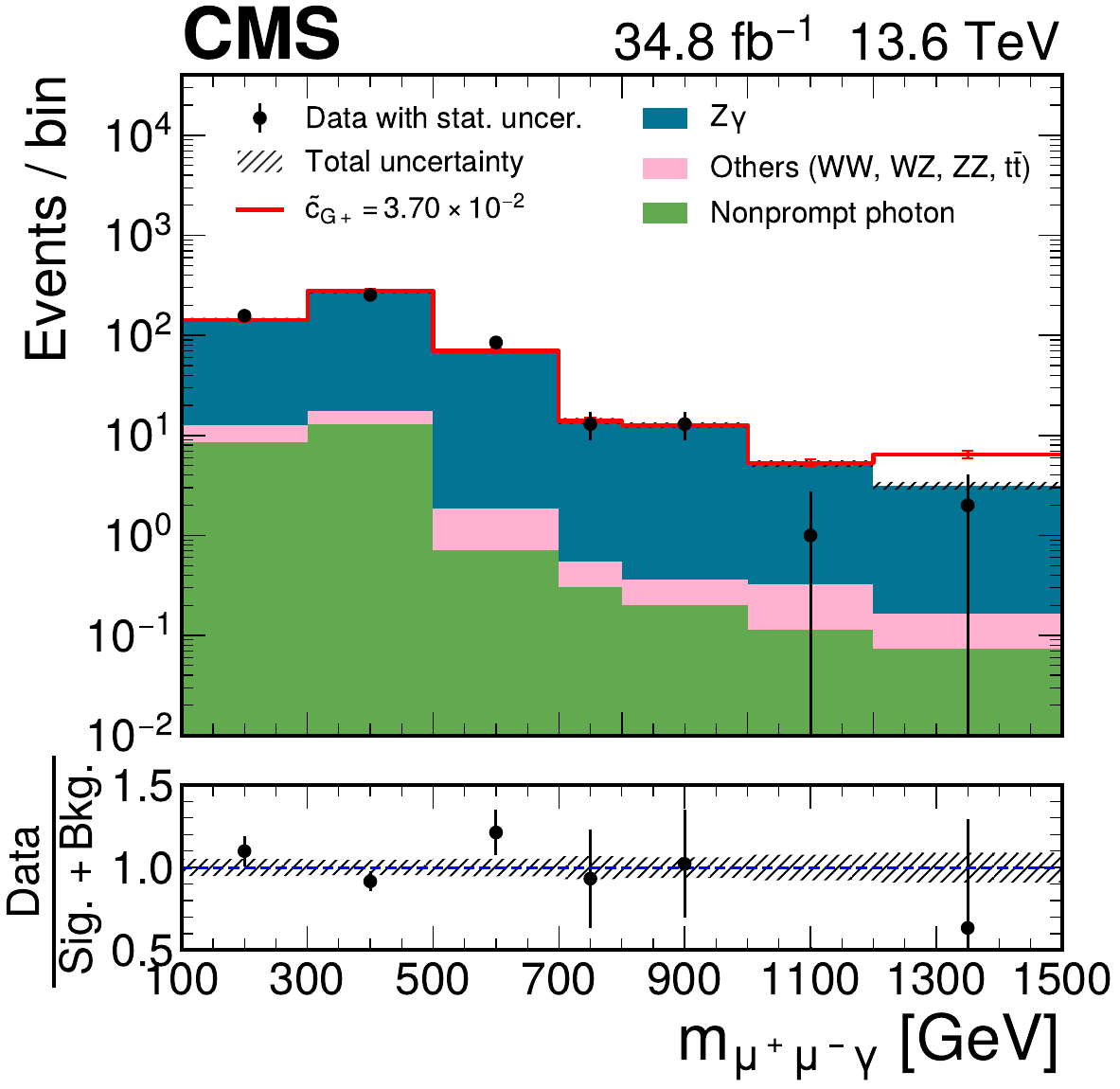}    
    \caption{Postfit distributions of $m_{\llg}$ for the electron channel (\cmsLeft) and the muon channel (\cmsRight) with the stricter photon \pt requirement. The experimental data are shown as points with associated statistical uncertainties, while the hatched bands represent the total uncertainties. Uneven bin sizes are used to ensure sufficient statistical precision in each bin. A representative nTGC signal \cGplus and its statistical uncertainty are shown as the red curve. The nTGC sensitivity mainly comes from the last bin which includes overflow events.}\label{fig:ntgc}
\end{figure}

The selection of BSM signal events follows the same criteria as in the fiducial cross section measurement study, except that the photon \pt requirement is increased to 120\GeV to be more sensitive to nTGC operators. The fit is performed in the $m_{\llg}$ range of 100--1500\GeV, in 7 bins. Uneven bin sizes $m_{\llg} = [100, 300, 500, 700, 800, 1000, 1200, 1500]\GeV$ are used to ensure sufficient statistical precision in each bin, the last bin includes the overflow. To obtain a continuous signal yield variation as a function of the coupling, the nTGC event yield over the SM event yield is described as a quadratic function of the coupling values for each $m_{\llg}$ bin. Expected and observed 95\% confidence level (CL) limits are computed for each operator coefficient, based on the profile likelihood ratio test statistic~\cite{CMS:2024onh}. Each operator coefficient is scanned independently with all other operator coefficients set to zero. Figure~\ref{fig:ntgc} shows the postfit $m_{\llg}$ distributions for the electron and muon channels. An example of the nTGC signal, with the new physics parameter \cGplus set to the value of the positive 95\% \CL limit observed in this study, is overlaid as a red curve. It is seen that most of the sensitivity to the nTGC signal comes from the highest $m_{\llg}$ bin.

Table~\ref{tab:nTGC_limits} shows the expected and observed 95\% \CL limits on the nTGC parameters. The results combine the measurements in the electron and muon channels. Two theoretical interpretations are considered: VPM and GSPM. Using Eq.~(\ref{eq:h-dim8}), the expected and observed limits in GSPM can be converted and compared with VPM, as shown in Table~\ref{tab:nTGC_limits_h4}. From the results of the converted limits, it can be seen that the limit on the vertex parameter $h_4$ differs between VPM and GSPM. The limits for GSPM, which maintains the full $\SUtwoL\otimes\UoneY$ gauge constrains, are significantly weaker than those obtained with VPM, which preserves only the unbroken \UoneEM symmetry. The expected (observed) limits differ by a factor of 46 (45). The factor by which the limits in the GSPM are weaker is consistent with that found in Ref.~\cite{Ellis:2022zdw}. The large differences observed between the two approaches highlight the strong model dependence in the extraction of anomalous gauge couplings, and may have important implications for the interpretation of future measurements. Tabulated results are provided in HEPData~\cite{HEPData}.

\begin{table}[htbp]
  \centering
  \topcaption{Expected and observed 95\% \CL limits on nTGC parameters for the combination of the measurements in the electron and muon channels. The first three rows show the results using VPM that preserves only the \UoneEM symmetry, while the last three rows show the results from GSPM that preserves the $\SUtwoL\otimes\UoneY$ symmetry.}
  { 
  \begin{scotch}{r r r r r}
    Limits & \multicolumn{2}{c}{Expected} & \multicolumn{2}{c}{Observed} \\
           & Lower           & Upper           & Lower           & Upper           \\
    \hline
    \hfourZ          & $-1.01\times10^{-6}$ & $1.01\times10^{-6}$ & $-8.17\times10^{-7}$ & $8.20\times10^{-7}$ \\
    \hthreeG      & $-8.34\times10^{-4}$ & $8.58\times10^{-4}$ & $-6.67\times10^{-4}$ & $6.88\times10^{-4}$ \\
    \hthreeZ           & $-7.01\times10^{-4}$ & $6.96\times10^{-4}$ & $-5.69\times10^{-4}$ & $5.48\times10^{-4}$ \\
    \cGplus       & $-3.83\times10^{-2}$ & $3.79\times10^{-2}$ & $-3.05\times10^{-2}$ & $3.06\times10^{-2}$ \\
    \cGminus       & $-3.09$              & $3.18$              & $-2.44$              & $2.54$              \\
    \cBW & $-1.24$           & $1.26$              & $-0.98$              & $1.00$             \\
  \end{scotch}
  }\label{tab:nTGC_limits}
  \end{table}

\begin{table}[htbp]
\topcaption{Expected and observed 95\% \CL limits for VPM parameter \hfourZ and for GSPM parameter \cGplus converted into \hfourZ via Eq.~(\ref{eq:h-dim8}). }
\centering
\begin{scotch}{l r r r r}
Limits & \multicolumn{2}{c}{Expected} & \multicolumn{2}{c}{Observed} \\
       & Lower & Upper & Lower & Upper \\
\hline
\hfourZ                     & $-1.01\times10^{-6}$ & $1.01\times10^{-6}$ 
                                & $-8.17\times10^{-7}$ & $8.20\times10^{-7}$ \\
$\hfourZ(\cGplus)$      & $-4.60\times10^{-5}$ & $4.64\times10^{-5}$ 
                                & $-3.71\times10^{-5}$ & $3.70\times10^{-5}$ \\
\end{scotch}\label{tab:nTGC_limits_h4}
\end{table}

In summary, a measurement of \Zg production using leptonic final states in proton-proton collisions at a center-of-mass energy of 13.6\TeV has been presented. The dataset used corresponds to an integrated luminosity of 34.8\fbinv, collected by the CMS experiment at the LHC during 2022.  Events with a pair of oppositely charged muons or electrons together with an isolated photon are selected. The measured fiducial cross section for the combined electron and muon channels is $1.896 \pm 0.033$ \stat $\pm 0.054 \syst \pm 0.006 \thy\unit{pb}$, in agreement with the standard model prediction of $1.922 \pm 0.094\unit{pb}$.  The $m_{\llg}$ distribution is used to set 95\% confidence level limits on neutral triple gauge couplings. These limits are derived, for the first time, in the context of a recently proposed framework of the effective field theory (EFT) based on dimension-8 operators, which maintains $\SUtwoL\otimes\UoneY$ gauge symmetry~\cite{Ellis:2022zdw}. Limits in the gauge symmetry preserving EFT are found to be more than an order of magnitude weaker for the affected parameter as predicted in reference~\cite{Ellis:2022zdw}.

\begin{acknowledgments}
We congratulate our colleagues in the CERN accelerator departments for the excellent performance of the LHC and thank the technical and administrative staffs at CERN and at other CMS institutes for their contributions to the success of the CMS effort. In addition, we gratefully acknowledge the computing centers and personnel of the Worldwide LHC Computing Grid and other centers for delivering so effectively the computing infrastructure essential to our analyses. Finally, we acknowledge the enduring support for the construction and operation of the LHC, the CMS detector, and the supporting computing infrastructure provided by the following funding agencies: SC (Armenia), BMBWF and FWF (Austria); FNRS and FWO (Belgium); CNPq, CAPES, FAPERJ, FAPERGS, and FAPESP (Brazil); MES and BNSF (Bulgaria); CERN; CAS, MoST, and NSFC (China); MINCIENCIAS (Colombia); MSES and CSF (Croatia); RIF (Cyprus); SENESCYT (Ecuador); ERC PRG, TARISTU24-TK10 and MoER TK202 (Estonia); Academy of Finland, MEC, and HIP (Finland); CEA and CNRS/IN2P3 (France); SRNSF (Georgia); BMFTR, DFG, and HGF (Germany); GSRI (Greece); NKFIH (Hungary); DAE and DST (India); IPM (Iran); SFI (Ireland); INFN (Italy); MSIT and NRF (Republic of Korea); MES (Latvia); LMTLT (Lithuania); MOE and UM (Malaysia); BUAP, CINVESTAV, CONACYT, LNS, SEP, and UASLP-FAI (Mexico); MOS (Montenegro); MBIE (New Zealand); PAEC (Pakistan); MES, NSC, and NAWA (Poland); FCT (Portugal);  MESTD (Serbia); MICIU/AEI and PCTI (Spain); MOSTR (Sri Lanka); Swiss Funding Agencies (Switzerland); MST (Taipei); MHESI (Thailand); TUBITAK and TENMAK (T\"{u}rkiye); NASU (Ukraine); STFC (United Kingdom); DOE and NSF (USA).

\end{acknowledgments}\section*{Data availability} Release and preservation of data used by the CMS Collaboration as the basis for publications is guided by the  \href{https://doi.org/10.7483/OPENDATA.CMS.1BNU.8V1W}{CMS data preservation, re-use and open access policy}.

\bibliography{auto_generated}\cleardoublepage \appendix\section{The CMS Collaboration \label{app:collab}}\begin{sloppypar}\hyphenpenalty=5000\widowpenalty=500\clubpenalty=5000\cmsinstitute{Yerevan Physics Institute, Yerevan, Armenia}
{\tolerance=6000
A.~Hayrapetyan, V.~Makarenko\cmsorcid{0000-0002-8406-8605}, A.~Tumasyan\cmsAuthorMark{1}\cmsorcid{0009-0000-0684-6742}
\par}
\cmsinstitute{Institut f\"{u}r Hochenergiephysik, Vienna, Austria}
{\tolerance=6000
W.~Adam\cmsorcid{0000-0001-9099-4341}, J.W.~Andrejkovic, L.~Benato\cmsorcid{0000-0001-5135-7489}, T.~Bergauer\cmsorcid{0000-0002-5786-0293}, M.~Dragicevic\cmsorcid{0000-0003-1967-6783}, C.~Giordano\cmsorcid{0000-0001-6317-2481}, P.S.~Hussain\cmsorcid{0000-0002-4825-5278}, M.~Jeitler\cmsAuthorMark{2}\cmsorcid{0000-0002-5141-9560}, N.~Krammer\cmsorcid{0000-0002-0548-0985}, A.~Li\cmsorcid{0000-0002-4547-116X}, D.~Liko\cmsorcid{0000-0002-3380-473X}, M.~Matthewman, I.~Mikulec\cmsorcid{0000-0003-0385-2746}, J.~Schieck\cmsAuthorMark{2}\cmsorcid{0000-0002-1058-8093}, D.~Schwarz\cmsorcid{0000-0002-3821-7331}, R.~Sch\"{o}fbeck\cmsAuthorMark{2}\cmsorcid{0000-0002-2332-8784}, M.~Shooshtari\cmsorcid{0009-0004-8882-4887}, M.~Sonawane\cmsorcid{0000-0003-0510-7010}, W.~Waltenberger\cmsorcid{0000-0002-6215-7228}, C.E.~Wulz\cmsAuthorMark{2}\cmsorcid{0000-0001-9226-5812}
% ...
\par}
$^{1}$Also at Yerevan State University, Yerevan, Armenia\\
$^{2}$Also at Technische Universit\"{a}t Wien, Vienna, Austria\\
% ...
\end{sloppypar}
\end{document}